\documentclass[prb,amsmath,amssymb,twocolumn,floatfix]{revtex4}
\usepackage{graphicx}
\usepackage{dcolumn}
\usepackage{bm}
\usepackage{subfigure}
\usepackage{amsmath}
\usepackage{feynmf}
\usepackage{hyperref}

\usepackage{attachfile}

\newcommand{\ra}{\rightarrow}

\newcommand{\bs}{\boldsymbol}

\usepackage{times}

\begin{document}
\title{Identifying the dominant pairing interaction in high-$T_c$ FeSe superconductors through Leggett modes}
\author{Wen Huang$^1$}
\author{Manfred Sigrist$^2$}
\author{Zheng-Yu Weng$^{1,3}$}
\affiliation{$^1$Institute for Advanced Study, Tsinghua University, Beijing, 100084, China \\
$^2$Institut f\"ur Theoretische Physik, ETH-Z\"urich, CH-8093 Z\"urich, Switzerland\\
$^3$ Collaborative Innovation Center of Quantum Matter, Tsinghua University, Beijing, 100084, China}

\date{\today}

\begin{abstract}
Heavily electron-doped and single-layer FeSe superconduct at much higher temperatures than bulk FeSe. There have been a number of proposals attempting to explain the origin of the enhanced transition temperature, including the proximity to magnetic, nematic and antiferro-orbital critical points, as well as possible strong interfacial phonon coupling in the case of single-layer FeSe. In this paper, we examine the effect of the various mechanisms in an effective two-band model. Within our model, the fluctuations associated with these instabilities contribute to different parts of the effective multiband interactions. We propose to use the collective phase fluctuation between the bands--the Leggett mode--as a tool to identify the dominant effective pairing interaction in these systems. The Leggett mode can be resolved by means of optical probes such as electronic Raman scattering. We point out that the Leggett mode in these systems, if present, shall manifest in the Raman $B_{1g}$ channel. 
\end{abstract}

\maketitle
\section{Introduction}
Since its inception in 2008 \cite{Kamihara:08}, the unconventional superconductivity in iron-based superconductors (FeSCs) has generated considerable interest. Besides superconductivity, the FeSCs exhibit a rich variety of electronic orders \cite{Paglione:10,Stewart:11,Dai:15}. Superconductivity in most FeSC families typically occurs in the vicinity of a stripe magnetic phase. The magnetic order usually follows a nematic transition at a slightly higher temperature \cite{Fang:08,Xu:08,Fernandes:11}. The nematic order parameter spontaneously breaks the four-fold rotational symmetry but preserves the translation symmetry of the underlying lattice. 

Despite enormous experimental and theoretical progresses, a unified understanding is still lacking regarding the superconducting mechanism in FeSCs \cite{Hirschfeld:11,Wang:11a,Chubukov:12,Fernandes:14,Si:16}. The proximity to the magnetic and nematic states indicates strong electron correlations and has led to a number of theories that attribute the superconducting pairing primarily to the magnetic \cite{Chubukov:08,Mazin:08,Kuroki:08}, orbital \cite{Kontani:10} and/or nematic fluctuations \cite{Lederer:15}, along with specific predictions for the order parameter symmetry and the gap structure on the multiple bands in the system. However, the complexity originating from the multiple strongly correlated Fe 3$d$-orbitals makes it difficult to unambiguously identify the primary mechanism(s) responsible for the formation of the various electronic orders. 

Among all the FeSC compounds, the FeSe family represents a notable outlier. Bulk FeSe superconducts below around $8$K at ambient pressure, but $T_c$ increases up to $37$K under pressure\cite{Medvedev:09} and can generically reach values above $30$K or even higher in heavily electron-doped FeSe.  In most cases, the electron-doping is achieved by means of intercalation \cite{Guo:10,Qian:11,Burrard:12,Lu:15}, such as in, e.g. K$_x$Fe$_{2-y}$Se$_2$ and (Li$_{0.8}$Fe$_{0.2}$)OHFeSe.  More strikingly, the single-layer FeSe epitaxially grown on SrTiO$_3$ and BaTiO$_3$ substrates shows a superconducting transition at temperatures well exceeding $50$K \cite{Wang:12,Peng:14a}, with indications of $T_c$ even above $100$K \cite{Ge:14}. The enhanced superconductivity in these high-$T_c$ FeSe compounds has sparked a substantial series of further investigations. 

Although bulk FeSe contains hole and electron Fermi pockets at the $\Gamma$- and $X/Y$-point, respectively, in the Brillouin zone (BZ) similar to other iron-pnictides, but with much smaller size of the Fermi surfaces \cite{Maletz:14,Terashima:14}. Moreover, the hole pocket is absent in the heavily electron-doped \cite{Qian:11,Niu:15,Zhao:16} and single-layer FeSe \cite{Liu:12,Peng:14}. The absence of a hole Fermi pocket at $\Gamma$, along with the drastically enhanced $T_c$, poses a serious challenge to the theories of spin-fluctuation-mediated superconducting pairing based on the quasi-nesting features between the electron and hole pockets \cite{Hirschfeld:11,Chubukov:08,Mazin:08,Kuroki:08}.

Similar to other FeSCs, the undoped FeSe exhibits a transition to nematic order at $90$K \cite{Bohmer:15}. However, in strong contrast to the former, the nematic transition is not followed by any long-range magnetic order down to the superconducting transition \cite{Baek:15}. Electron doping suppresses the nematic order, while magnetism continues to be absent up to the optimal doping level \cite{Wen:16}. Nevertheless, inelastic neutron scattering studies on undoped FeSe reported pronounced spin fluctuations \cite{Wang:16}, and the standard stripe magnetic order common to other FeSCs does emerge under applied pressure \cite{Terashima:15,Sun:16,Wang:16b,Kothapalli:16}. In addition, rich spin excitation spectra are commonly observed in alkali-metal intercalated \cite{Taylor:12,Friemel:12,Park:11} as well as alkali-hydroxy-intercalated \cite{Pan:16,Ma:17} electron-doped FeSe-compounds. Furthermore, despite the lack of definitive experimental evidence to date, it is sensible to also pay attention to the antiferro-orbital (AFO) ordering associated with the degenerate and strongly correlated $d_{xz}$ and $d_{yz}$-orbitals. 

The generically enhanced superconducting pairing in the high $T_c$ FeSe compounds seems to connect closely with their peculiar electronic structure with only electron pockets, dichotomy of nematic and magnetic ordering, possible AFO ordering and the presence of their fluctuations, as well as the substrate environment in the case of single-layer FeSe. This naturally motivates the question as to how the nematic, spin and AFO fluctuations may cooperate with the unique electronic structure to strengthen the effective pairing interactions, and how $T_c$ seems to increase further in the presence of interfacial phonon coupling in single-layer FeSe \cite{Lee:15,Lee:14}. 

In this paper, we do not aim to provide a microscopic theory behind the enhancement of superconductivity as in some recent theoretical works \cite{Xiang:12,Rademaker:15,Dumitrescu:16,Yamakawa:16a,Yamakawa:16b}. Instead, similar to Li {\it et al.} \cite{Li:16}, we take an effective two-band model and assume a priori the presence of various fluctuation and/or phonon-induced interactions within and between the bands, i.e. intra- and interband interactions. Notably, nematic fluctuations and the interfacial phonon coupling mainly contribute to intraband interaction, while particular types of spin and AFO fluctuations at wavevector $(\pi,\pi)$ dominate the interband interaction. When the intraband exceeds the interband interaction, the superconducting state shall exhibit a well-defined collective phase mode -- the so-called Leggett mode \cite{Leggett:66} -- which corresponds to the relative phase fluctuations between the two bands. We propose that the existence or nonexistence and the characteristic energy of the Leggett mode if present, can help to elucidate the relative strength of the various contributions to the pairing interactions. The Leggett mode can be probed in optical measurements such as Raman scattering \cite{Blumberg:07}. Note that the Leggett modes in the respective scenarios with dominant intra- and interband interactions have been discussed earlier in a general context \cite{Marciani:13,Cea:16}, while in the present study we focus on heavily-electron doped and single-layer FeSe systems which exhibit distinct Fermi surface geometry. Furthermore, the collective modes -- including Leggett modes -- in iron-arsensic superconductors have also been thoroughly discussed in recent years \cite{Lin:12,Ota:11,Khodas:14,Maiti:15,Maiti:16,Maiti:17a}. 

\begin{figure}
\includegraphics[width=5cm]{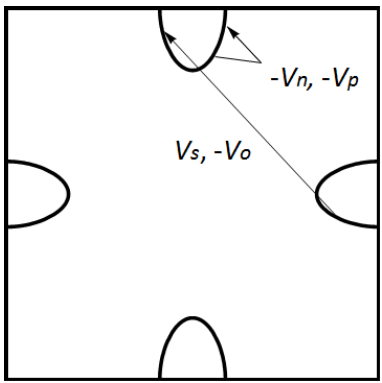}
\caption{Sketch of the Fermi surfaces of heavily electron-doped and single-layer FeSe. The primary intra- and interband interactions considered in our model are shown, including the respective SDW-fluctuation and AFO-fluctuation mediated interband interactions $V_s$ and $-V_o$, the nematic-fluctuation induced intraband interaction $-V_n$,  and the phonon induced intraband interaction $-V_p$ for single-layer FeSe.}
\label{fig:FS}
\end{figure}

We also remark that, parallel to the two-band description adopted here, some theories explored the possible crucial role of the incipient hole band at the $\Gamma$-point in the BZ. The hole band may develop an incipient Cooper pairing induced by the magnetic fluctuations associated with either the local moments \cite{You:11} or the itinerant carriers \cite{Linscheid:16}. On this basis, You {\it et al.} \cite{You:11} further noted that, since the quasi-nesting between the electron and hole bands is suppressed, superconductivity in high-$T_c$ FeSe systems may have benefited from the absence of a competing itinerant spin-density wave order. \\


\section{ Effective models}
We use a two-band model with two electron pockets at the $X/Y$-points of the single-Fe BZ to mimic the electronic band structure in heavily electron-doped FeSe, as in Fig.\ref{fig:FS}. The interactions between the low-energy electrons should, in principle, be sensitive to the microscopic details such as the orbital composition at the Fermi level. However, since we are concerned with the properties arising from the couplings between the individual superconducting bands, we may disregard the details of the momentum-space structure of the intra- and interband interactions, but rather take a simplified form for the {\it integrated} Cooper channel effective interactions within and between the bands. 



We begin by writing down the effective action, 
\begin{equation}
S= S_{el} + \sum_{\nu=s,o,n,p}(S_{\nu} + S_{el-\nu}) \,
\end{equation}
where $S_{el}$ denotes the action of the two itinerant electron bands located around the $X$ and $Y$-points; $S_\nu$ is the action of the bosonic mode $\phi_\nu$ where $\nu = s,o,n,p$ stand for SDW, AFO, nematic and phonon modes, respectively; and $S_{el-\nu}$ describes the Yukawa-type coupling between the electrons and the bosonic modes to be discussed in turn below.   


As in other FeSC's, spin fluctuations are universally present in FeSe compounds. In undoped bulk FeSe, neutron scattering reveals coexisting N\'eel spin fluctuations at $(\pi,\pi)$ and stripe spin fluctuations at $(\pi,0)$ \cite{Wang:16}. Furthermore, heavily electron doped $A_x$Fe$_{2-y}$Se$_2$ \cite{Taylor:12,Friemel:12,Park:11}, and more recently Li$_{0.8}$Fe$_{0.2}$ODFeSe~\cite{Pan:16,Ma:17} were found to exhibit spin resonant excitations at wavevectors surrounding $(\pi,\pi)$. The $(\pi,0)$ fluctuations are ineffective in mediating Cooper pairing in our model due to the lack of a hole pocket at the $\Gamma$-point. The fluctuations at wavevector $(\pi,\pi)$ (or wavevectors that connect the nearly nested portions of the two Fermi pockets \cite{Friemel:12,Pan:16}, same below), on the other hand, actively scatter electrons between the two pockets and should, therefore, play an important role. The coupling between the fluctuating SDW field and the itinerant carriers reads,
\begin{equation}
\mathcal{S}_{el-s} = \lambda_s\int d\vec{q} \vec{\phi}_s (\vec{q})\cdot \sum_{\alpha,\beta}\int d\vec{k} c^\dagger_{\vec{k}+\vec{q},\alpha}\vec{\sigma}_{\alpha\beta} c_{\vec{k},\beta} + h.c. \,. 
\label{eq:Lphis}
\end{equation}
Here the components of $\vec{\sigma}$ are the Pauli matrices. Integrating out the SDW field returns a spin-dependent effective interaction peaking at momentum transfer $(\pi,\pi)$, i.e. $V_s(\vec{k},\vec{p}) \propto -\chi_s(\vec{k}-\vec{p})\vec{\sigma}_{\alpha\beta}\cdot \vec{\sigma}_{\gamma\delta}$, where $\chi_s$ stands for the SDW magnetic susceptibility. This wavevector connects the two Fermi pockets, hence the interaction is predominantly interband. Such an effective interaction in the singlet pairing channel amounts to a repulsive interaction peaking at the same momentum transfer, thereby promoting sign-changing superconducting gaps on the two bands,  i.e. a node-less $d$-wave pairing. 

Likewise, AFO fluctuations in FeSCs may also develop predominantly at wavevectors $(\pi,\pi)$ and $(\pi,0)$. We consider the former wavevector, for which the fluctuations scatter electrons between the two pockets. There is, however, an important distinction from the SDW fluctuations, in that here the scattering is spin-independent, 
\begin{equation}
\mathcal{S}_{el-o} = \lambda_o\int d\vec{q} \phi_o (\vec{q}) \sum_{\sigma}\int d\vec{k} c^\dagger_{\vec{k}+\vec{q},\sigma} c_{\vec{k},\sigma} + h.c. \,. 
\label{eq:LphiAFO}
\end{equation}
This leads to an effective interaction in the Cooper channel, $V_o(\vec{k},\vec{p})  \propto -\chi_o(\vec{k}-\vec{p})$, which is primarily attractive and interband, thus favoring a sign-conserving $s$-wave pairing. As a consequence, the AFO fluctuation and the SDW fluctuation mentioned above compete against each other. 

The nematic susceptibility $\chi_n$, whether spin- or orbital-driven, peaks at zero momentum. Hence nematic fluctuations are effective in scattering electrons by small momenta, in a manner given by the effective action, 
\begin{equation}
\mathcal{S}_{el-n} =\lambda_n\int \vec{q} \phi_n (\vec{q})\sum_{\sigma} \int d\vec{k} c^\dagger_{\vec{k}+\vec{q},\sigma} c_{\vec{k},\sigma} + h.c. \,. 
\label{eq:Lphin}
\end{equation}
This scattering is also spin-independent. The induced effective electron interaction, $V_n(\vec{k},\vec{p}) \propto -\chi_n(\vec{k}-\vec{p})$, is attractive and predominantly intraband. This interaction alone drives electron pairing, and should give rise to degenerate sign-changing ($d$-wave) and sign-conserving ($s$-wave) gaps. The degeneracy can be lifted by either interband interactions or interband hybridization, the latter of which has been discussed in Ref. \onlinecite{Kang:16}. 

Taking together, the primary multiband interactions can be expressed in the matrix form as,
\begin{equation}
\hat{V}=\begin{pmatrix}
-V_n  &  V_s-V_o \\
V_s-V_o   &  -V_n 
\end{pmatrix} \,,
\label{eq:V}
\end{equation}
where we take $V_n,V_s,V_o>0$ for notational clarity. Solving a coupled BCS gap equation using (\ref{eq:V}) yields gap functions on the two bands, the more attractive of which characterizes the stable ground state. Since the two bands have the same density of states, the gap functions are equivalent to the eigenvectors of $\hat{V}$. The solution thus obtained denotes the relative sign and magnitude of the band gaps. The preference between sign-changing and sign-conserving pairings is determined by relative strength of the two interband interactions, i.e. the former is favored if $V_s>V_o$, otherwise the latter is more stable. 

A cooperative mechanism of a certain subset of the multiband interactions may be crucial for the boost of $T_c$. In particular, for the leading superconducting solution the intra- and net interband interactions do not compete: since the most negative eigenvalue of $\hat{V}$ is given by $-V_n-|V_s-V_o|$, both intra- and interband interactions act to strengthen pairing. The enhancement is most effective when either $V_s$ or $V_o$ dominates the interband interaction. 


In light of the striking observation of a replica band in single-layer FeSe suggestive of a strong small-momentum scattering by the interfacial phonons \cite{Lee:14}, phonon-induced effects should be properly accounted for in this system. This gives an effective action analogous to the one formulated for nematic fluctuations (\ref{eq:Lphin}). As a consequence, the phonon coupling gives rise to an attractive interaction $-V_p$ $(V_p>0)$ in the same fashion as the nematic fluctuations mediating $-V_n$. Hence the total effective interaction becomes, 
\begin{equation}
\hat{V}=\begin{pmatrix}
-V_n-V_p  &  V_s-V_o \\
V_s-V_o   &  -V_n-V_p 
\end{pmatrix} \,. 
\label{eq:V1}
\end{equation}
The effective pairing interaction is then given by $-V_n-V_p-|V_s-V_o|$, from which it is easy to see the conducive role of the interfacial phonons in enhancing superconductivity. 

In the following, we first discuss the existence/nonexistence of Leggett mode in the presence of various dominant pairing interactions, and then proceed to discuss the detection of the Leggett mode when it does exist. \\

\section{Leggett mode energy} 
As was originally predicted by Leggett \cite{Leggett:66}, the interband interaction gives rise to an effective Josephson coupling between the superconducting order parameters on different bands, which locks their relative phase. Under external perturbations, the relative phase can oscillate in time, costing a finite amount of energy that is determined by the interband coupling. This is the {\it Leggett mode}. As Leggett pointed out in his work (in the context of a two-band $s$-wave superconductor), this mode corresponds essentially to a collective fluctuation between the leading and subleading superconducting states \cite{footnote}. To this end, we note another collective superconducting mode, the so-called Bardasis-Schrieffer (BS) mode \cite{Bardasis:61}, which has been discussed previously in the context of iron-pnictide superconductors \cite{Khodas:14,Maiti:15,Maiti:16,Maiti:17a}. Usually, this mode describes fluctuations between leading and subleading states in distinct Cooper pair angular momentum channels, and can also exist in single-band systems. As one can see, the collective mode we study here can be regarded both as a Leggett and a BS mode. 


Naturally, the characteristic energy of the Leggett mode encodes crucial information about the multiband interactions in FeSe systems.  Below we analyze the Leggett modes in the presence of various configurations of multiband interactions. The expressions of the Leggett mode energy is derived in Appendix \ref{appA}.

We first ignore the phonon contribution $V_p$. Of particular interest are the two limiting cases where the pairing is driven primarily by the nematic or by SDW/AFO fluctuations. In the former, $V_n\gg |V_s-V_o|$, the interband Josephson coupling reads $J=|V_s-V_o|/(V_n^2-|V_s-V_o|^2) >0$ (see Appendix \ref{appA}). Both solutions to the gap equation correspond to attractive superconducting channels, i.e., leading $d$-wave with subleading $s$-wave pairing if $V_s>V_o$, or vice versa if $V_s<V_o$. Thus, a coherent Leggett mode exists, and its resonance energy is given by, 
\begin{equation}
w_L = \sqrt{2\frac{J}{N_0}} \Delta_0 =  \sqrt{\frac{2}{\lambda}}\sqrt{\frac{|V_s-V_o|}{V_n-|V_s-V_o|}} \Delta_0 \,,
\label{eq:L1}
\end{equation}
where $\Delta_0$ is the superconducting gap, $N_0$ is the density of states of a single band, and $\lambda= N_0 (V_n+|V_s-V_o|)$ gives the effective coupling strength in the leading superconducting channel. Taking a rough approximation $\lambda \sim 1$, $w_L \sim \sqrt{2|V_s-V_o|/V_n}\Delta_0 \ll 2 \Delta_0$, which is much smaller than the quasiparticle continuum edge at $w=2\Delta_0$, consistent with Ref. \onlinecite{Cea:16}. Such a soft mode reflects the near-degeneracy between the leading and subleading pairing states, and equivalently the relative ease in fluctuating the relative phase between the two bands, when the interband interaction is weak. 

In the other limit where the interband interaction dominates, i.e. $V_n < |V_s-V_o|$, no coherent Leggett mode is present. This can be understood as follows. When the the strength of the interband interaction exceed that of the intraband one, there is only one attractive superconducting solution. As a result, no subleading superconducting state exists to be excited to.

In the intermediate regime with $|V_s-V_o|  \lesssim V_n$, $w_L$ is again given by (\ref{eq:L1}). The characteristic energy $w_L$ increases with growing $|V_s-V_o|/V_n$, exceeding the continuum edge for $|V_s-V_o|\sim V_n/2$. In this regard, $w_L$ is a qualitative measure of the relative strength between the intra- and interband interactions.  Finally, in accordance with the discussions in the previous section, the phonons strengthen intraband interactions. Therefore, the addition of $V_p$ broadens the parameter range where a coherent Leggett mode can appear.\\


\section{Detection of Leggett mode}
The Leggett modes couple to electromagnetic fields and, hence, can be excited by photons in optical measurements, such as the electronic Raman scattering, as has been demonstrated for the two-band superconductor MgB$_2$ \cite{Blumberg:07}. They manifest as resonance features in the Raman spectrum when the difference between the frequencies of the incident and scattered photons matches that of the Leggett modes in an appropriate scattering channel. Note when $w_L$ exceeds the continuum edge, the Leggett mode becomes damped by quasiparticles excitations \cite{Cea:16}, and the broadened resonance peak overlaps with the quasiparticle continuum spectrum at $w>2|\Delta|$, wherein the measured $Q$-factor may be small (as is the case in MgB$_2$ \cite{Blumberg:07}). 

\begin{figure}
\subfigure{\includegraphics[width=4cm]{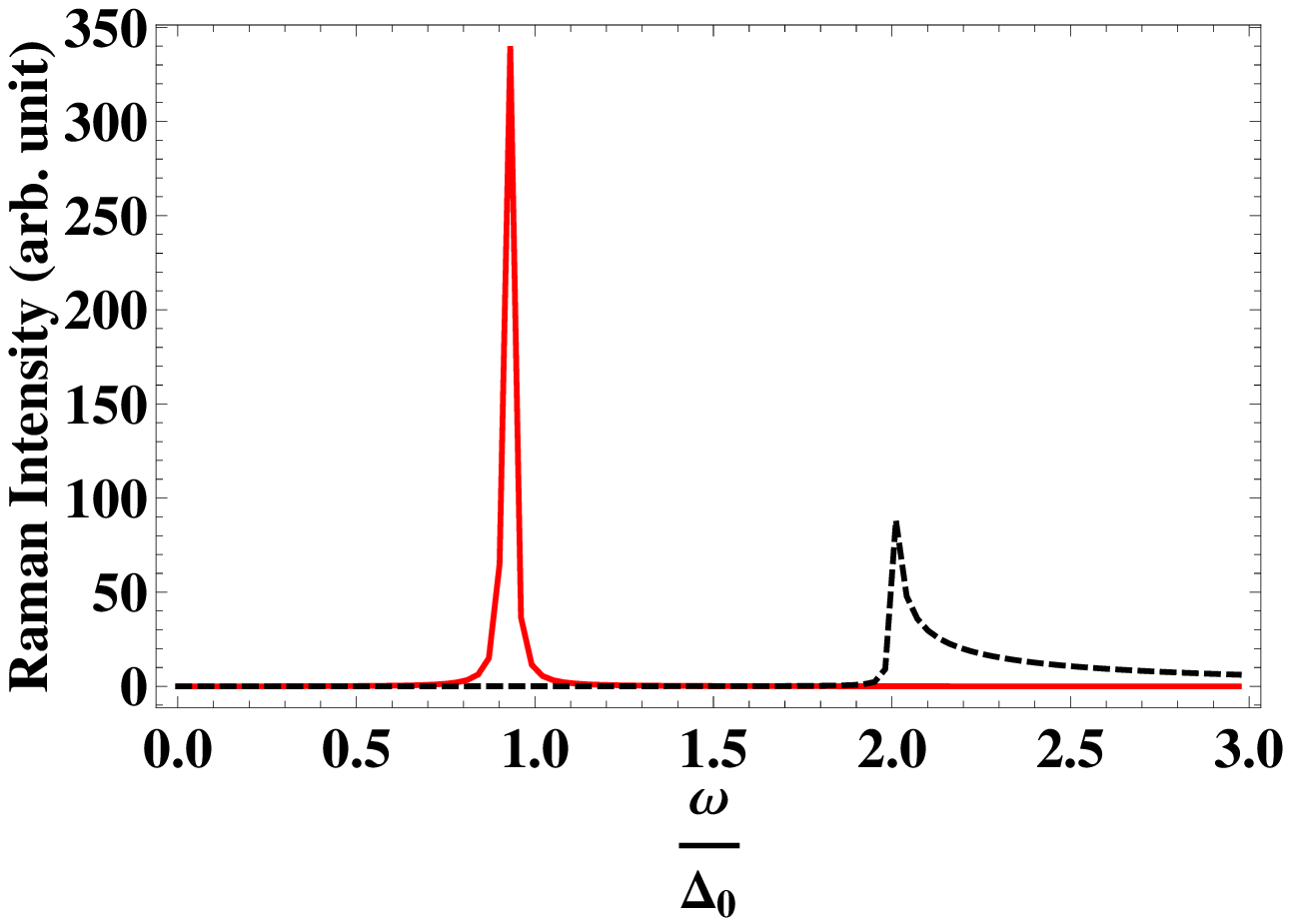} }
\subfigure{\includegraphics[width=4cm]{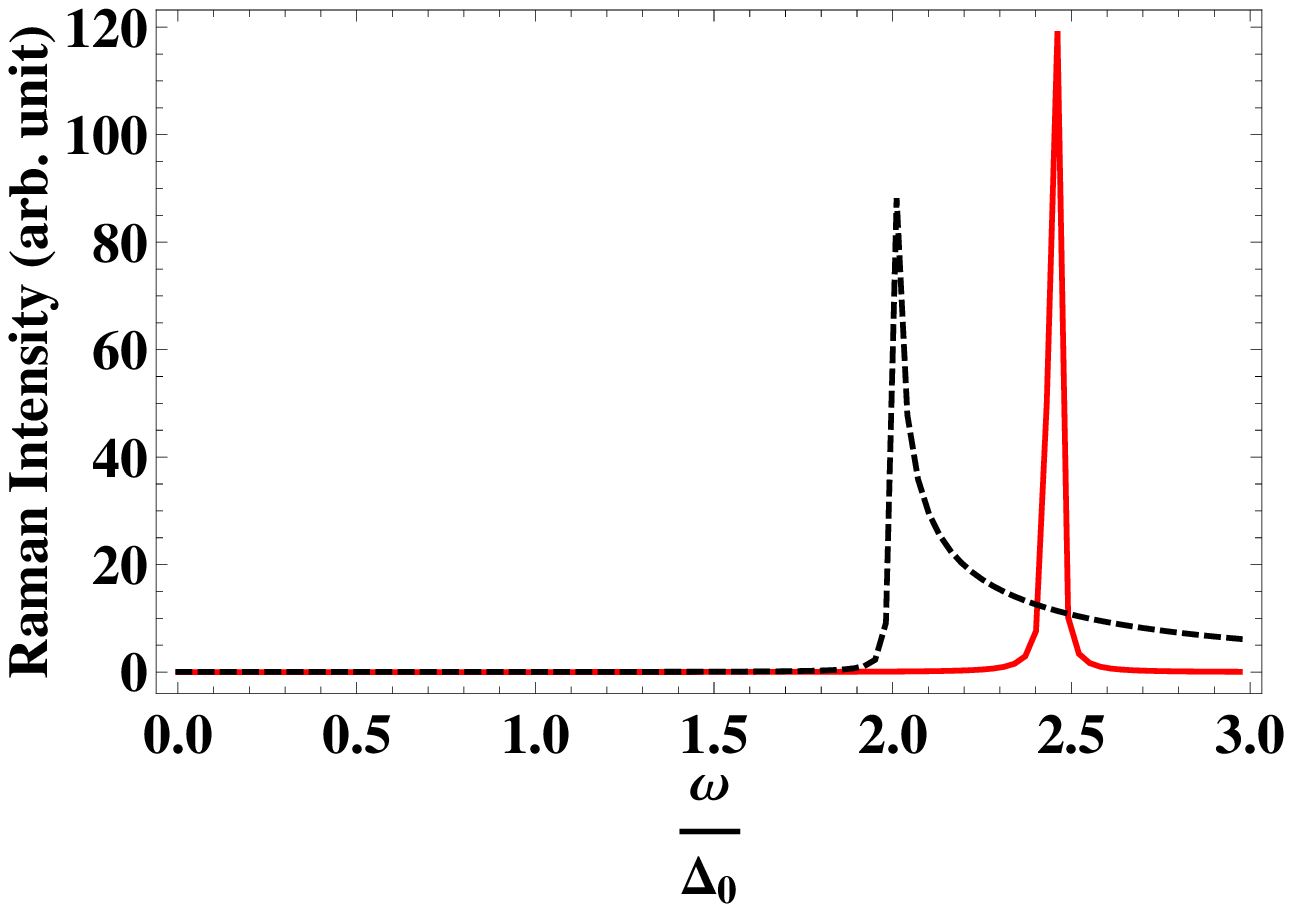} }
\caption{Schematic Raman intensity in the $B_{1g}$ channel for two models with the ratio of inter- and intraband interactions $0.3$ (left) and $0.75$ (right). In both cases the intraband interaction is stronger than the interband one. The continuum contribution (black dashed) shows a peak at $w=2\Delta_0$, while the Leggett resonance (red solid) occurs at $w_L$. We have assumed an isotropic superconducting gap on the two bands, taking the approximation $\lambda \sim1$ in (\ref{eq:L1}), and used a small imaginary component $\tau=0.002\Delta_0$ to yield broadened peaks. The formula used for these calculations are derived in Appendix \ref{appB}. On the right figure we have ignored the quasiparticle damping of the Leggett mode peak.}
\label{fig:Raman}
\end{figure}

Traditionally, the Leggett mode arises in the Raman $A_{1g}$ channel, as the associated leading and subleading superconducting states belong with the same Cooper pair angular momentum channel. However, in heavily electron-doped and single-layer FeSe, the Leggett mode amounts to a collective fluctuation between $s$- and $d$-wave states. This originates from the unique Fermi surface topology with the two Fermi pockets locating around the $X$ and $Y$-points in the BZ (Fig. 1). Consequently, the Leggett mode under consideration will emerge in the Raman $B_{1g}$ channel. In particular, as is shown more explicitly in Appendix \ref{appB}, the coupling between the collective fluctuation and the Raman vertex is proportional to the following mean value,
\begin{equation}
\langle f_{0\vec{k}} f_{1\vec{k}} \gamma_{\vec{k}} \rangle_\text{FS} \,,
\label{eq:criterion}
\end{equation}
where $\langle \cdots \rangle_\text{FS}$ represents an average over the Fermi surface, $f_{s\vec{k}}$ and $f_{d\vec{k}}$ are character functions of the $s$ and $d$-wave symmetry, and $\gamma_{\vec{k}}$ is the Raman vertex in a particular channel. It is immediately clear that by symmetry only the $B_{1g}$-channel yields a nonvanishing coupling. Figure 2 shows two representative schematic $B_{1g}$ Raman spectra when the intraband interaction plays the leading role in driving the pairing.

Related Raman scattering measurements have been performed on the heavily electron-doped intercalated compound, Rb$_{0.8}$Fe$_{1.6}$Se$_2$ \cite{Kretzschmar:13}. There, below the continuum edge, apart from a phonon mode, no additional peak was visible in the $B_{1g}$ channel with features that could be associated with a Leggett resonance. It is also unclear whether any resonance is present in the $B_{1g}$ spectrum above the continuum edge. Hence we cannot conclude on the qualitative comparison between the intra- and interband interactions. Nonetheless, following our arguments above, the absence of such a resonance below $2\Delta_0$ suggests that the interband interaction at least constitutes a non-negligible ingredient of the total effective pairing strength in this particular compound. Corroborating the significance of interband interactions, the superconducting magnetic resonant modes observed in neutron scattering in several heavily electron-doped intercalated compounds \cite{Taylor:12,Friemel:12,Park:11}, including Rb$_{0.8}$Fe$_{1.6}$Se$_2$ \cite{Friemel:12}, appeared at wavevectors which most likely connect the two bands\cite{Friemel:12,Pan:16}.

Assisted by the strong coupling to interfacial phonons \cite{Lee:14}, the intraband interaction is expected to be boosted significantly in the single-layer FeSe grown on $A$TiO$_3$ substrates. A coherent Leggett mode is thus more likely to emerge in these systems. However, due to the finite optical penetration into the substrate, the Raman spectroscopy may see a much stronger background noise signal, making it difficult to disentangle the authentic response of the FeSe layer. It is, thus, necessary to devise a careful Raman measurement to search for such a Leggett resonance there.  \\

\section{Conclusions}
In this paper, we outline the possible main sources of the multiband interactions in the two-band high-$T_c$ heavily electron-doped and single-layer FeSe superconductors. The nematic fluctuations and/or interfacial phonons contribute primarily to the intraband interaction, while the SDW and AFO fluctuations at momentum $(\pi,\pi)$ (or similar wavevectors connecting the two pockets) mainly drive competing interband interactions. If the net interband interaction is weaker than the intraband one, a novel collective phase excitation--a Leggett mode--shall arise. We propose that optical probes such as Raman spectroscopy can provide crucial information regarding the relative strength of the various contributions to the effective pairing glue.

\acknowledgements
We are grateful to Hong Yao for many valuable discussions, to Saurabh Maiti for critical comments on our manuscript and to Qingming Zhang for a helpful communication. ZYW acknowledges support from NSFC and MOST of China. WH is supported by the C.N. Yang Junior Fellowship at the Institute for Advanced Study at Tsinghua University and is grateful for the hospitality of the Pauli Center for Theoretical Physics at ETH Zurich.

\pagebreak
\widetext
\setcounter{equation}{0}
\setcounter{figure}{0}
\setcounter{table}{0}
\setcounter{page}{1}
\makeatletter
\renewcommand{\thefigure}{A\arabic{figure}}

\appendix
\section{Derivation of Leggett mode energy}
\label{appA}
Here, we derive the characteristic energy of the Leggett modes in our effective two-band model introduced in the main text. We take the effective interaction (\ref{eq:V}) in the main text for illustration. For simplicity, we include below only the nematic and spin fluctuation mediated interactions, $V_n$ and $V_s$. The other interactions can be accounted for easily in an analogous fashion. The coupled BCS gap equation is written as, 
\begin{equation}
\begin{pmatrix}
\Delta_1 \\
\Delta_2
\end{pmatrix} = - N_0 \text{ln}\left( \frac{\pi W_c}{2 T_c}\right) 
\begin{pmatrix}
-V_n  & V_s \\
V_s & -V_n
\end{pmatrix}
\begin{pmatrix}
\Delta_1 \\
\Delta_2
\end{pmatrix} \,,
\end{equation}
where $N_0$ is the density of states of a single band, $T_c$ is the superconducting transition temperature, $W_c$ is some characteristic cutoff energy, and we have chosen $V_n,V_s>0$ for notational clarity. In principle, the cutoff should be different for interactions mediated by nematic and magnetic fluctuations, but for simplicity we take it to be the same for both. Solving the gap equation is equivalent to diagonalizing $\hat{V}$, the latter of which returns the effective pairing interaction in the eigen-channels: $-V_n-V_s$ and $-V_n+V_s$. The two eigenvectors are $(1,-1)$ and $(1,1)$. The first solution corresponds to the superconducting ground state, which describes a sign-changing node-less $d$-wave pairing. Note if the attractive interband interaction $-V_o$ is included and if $V_o>V_s$, the second solution, i.e. the sign-preserving $s$-wave state, is favored. 

The effective Josephson coupling between the bands can be captured in the following effective action \cite{Ota:11,Lin:12,Huang:16}, 
\begin{equation}
\mathcal{S}_\Delta = \int \hat{\Delta}^\dagger (-\hat{V}^{-1}) \hat{\Delta} -\frac{1}{\beta}\sum_{l=1,2} \text{Tr~ln}G^{-1}_l   \,,
\label{eq:Josephson}
\end{equation}
where $\hat{\Delta} = (\Delta_1,\Delta_2)^T$ denotes the superconducting order parameter on the two bands, $\beta=1/T$ with $T$ the temperature, and the $l$-band Gor'kov Green's function is given by 
\begin{equation}
\hat{G}^{-1}_l= -\begin{pmatrix}
\partial_\tau -\frac{\nabla^2}{2m}-\mu &  \Delta_l \\
\Delta_l^\ast & \partial_\tau + \frac{\nabla^2}{2m}+\mu
\end{pmatrix}  \,. 
\label{eq:Greenfunction}
\end{equation} 
In (\ref{eq:Josephson}), we have assumed that $\hat{V}^{-1}$ is non-singular, i.e. $\text{det}\hat{V}$ is non-vanishing. The interband Josephson coupling $J$ is defined by the off-diagonal element of the inverse of -$\hat{V}$,
\begin{equation}
J = \frac{V_s}{\text{det}{\hat{V}}} \,,
\label{eq:Josephson1}
\end{equation} 
Considering now small deviations of the $U(1)$ phase of the order parameter from the stable state, $\theta_l=\theta_{0l}+\phi_l$ $(l=1,2)$, the action in Eq.({\ref{eq:Josephson}) can be expanded with respect to the $\phi_l$'s. In particular, the first term of ({\ref{eq:Josephson}) gives the following Josephson coupling terms,
\begin{eqnarray}
&&J(\Delta_1^\ast\Delta_2 + \Delta_2^\ast\Delta_1)\nonumber \\
&=& 2J|\Delta|^2 \cos (\theta_1-\theta_2) \nonumber \\
&=& 2J|\Delta|^2[\cos(\theta_{01}-\theta_{02})\cos(\phi_1-\phi_2) -\sin(\theta_{01}-\theta_{02})\sin(\phi_{1}-\phi_{2}) ] \nonumber \\
&=& -J|\Delta|^2\cos(\theta_{01}-\theta_{02}) (\phi_1 - \phi_2)^2 + ... \,.
\label{eq:Exp1}
\end{eqnarray}
In the last equation `...' stands for the sum of an unimportant constant and higher order terms $\mathcal{O}(\phi^4)$. Note that linear terms in $\phi$ do not survive, as $\sin(\theta_{01}-\theta_{02})$ vanishes for the superconducting solutions because $\theta_{01}-\theta_{02}=n\pi$ where $n$ is an integer. 

A gradient expansion of the second term of (\ref{eq:Josephson}) with respect to $\phi_l$ returns the Goldstone mode action associated with the individual bands \cite{Altland:10},
\begin{equation}
- \frac{1}{\beta}\sum_{l=1,2} \text{Tr~ln}G^{-1}_l =\sum_{l=1,2} \left[ N_0(\partial_\tau \phi_l )^2 + \frac{N_0\bar{v}_F^2}{2} (\bs\nabla\phi_l)^2 \right] \,,
\label{eq:Exp2}
\end{equation} 
where $N_0$ denotes the single-band density of states at the Fermi energy and $\bar{v}_F$ is the average Fermi velocity.

Performing Fourier transformation of (\ref{eq:Exp1}) and (\ref{eq:Exp2}), the effective action for $\phi_l$'s become,
\begin{equation} 
\mathcal{S}[\phi] =\int dq  \hat{\phi}^T(-q) \mathcal{M} \hat{\phi}(q) \,,
\label{eq:phaseAction}
\end{equation}
where $q=(q_0,\vec{q})$ with $q_0$ being the bosonic Matsubara frequency, and $\hat{\phi}=(\phi_1,\phi_2)^T$, and the matrix,
\begin{equation}
\mathcal{M} = 
\begin{pmatrix}
\mathcal{K}-J \epsilon_{12} & J \epsilon_{12} \\
J\epsilon_{12} & \mathcal{K}-J \epsilon_{12}
\end{pmatrix} 
\label{eq:M}
\end{equation}
where $\mathcal{K} = N_0 (q_0^2 + \bar{v}^2_F q^2/2)$, and $\epsilon_{12}= \cos(\theta_{01}-\theta_{02})|\Delta|^2$. Since the repulsive $V_s$ favors sign-changing pairing, $\epsilon_{12}\equiv -|\Delta|^2$ in our model. After an analytic continuation by replacing $iq_0 \ra w + i0^+$, the dispersion relations for the phase modes may be obtained by diagonalization (\ref{eq:M}),
\begin{eqnarray}
w^2_G &=& \frac{1}{2} \bar{v}^2_{F}q^2 \,,\\
w^2_{L}&=& 2\frac{|\Delta|^2}{N_0}J+  \frac{1}{2}\bar{v}^2_{F}q^2 \,.
\label{eq:dispersion}
\end{eqnarray}
Here $w_G$ denotes the usual gapless $U(1)$ Goldstone mode, which would be massive had we properly included the vector potential in our formalism; $w_{L}$ is the Leggett mode, whose excitation gap is determined by the interband Josephson coupling. Note that if $J<0$, as would be the case for $V_s>V_n$, no physical solution exists for $w_L$, i.e. the Leggett mode is overdamped. This peculiar scenario corresponds to the absence of coherent relative phase oscillations when the pairing is overly dominated by the interband interaction \cite{Marciani:13}. To understand this, first note that the Leggett mode corresponds essentially to a fluctuation between the leading and subleading pairing channels mentioned above ($d$- and $s$-waves in our case). If the subleading channel becomes repulsive, i.e. $-V_n+V_s>0$, there exists no true subdominant superconducting channel for the ground state to coherently excite to. As a consequence, a coherent Leggett mode cannot exist in this scenario. 

\section{Derivation of Raman response function}
\label{appB}
This section presents a derivation of the Raman response function used for calculations in the main text. We first note that the collective Leggett mode studied in the main text is simultaneously a Bardarsis-Schrieffer (BS) mode, i.e. the relative phase oscillation between the $X$- and $Y$-pockets correspond exactly to a fluctuation between $s$- and $d$-wave channels, if we view the system as an effective one-band model. It will be seen that the corresponding fluctuation couples to the Raman $B_{1g}$-channel. To make the symmetry argument transparent, we will adopt the BS description. 

Denoting the ground state pairing $\Delta_0 f_{0\bs k}$ and the subleading pairing $\psi f_{1\bs k}$ where $f_{0\bs k}$ and $f_{1\bs k}$ are form factors characteristic of the corresponding Cooper pair angular momentum channels. In the previous section we show that the Leggett (BS) mode has excitation energy $w_L$,  the effective action of the collective fluctuations of $\psi$ alone can then be written as,
\begin{equation}
S[\psi] = \int d\tau d\bs r \psi^\ast(\tau,\bs r) (-\partial_\tau^2  -\frac{\bar{v}^2_F}{2}\bs \nabla^2+w_L^2) \psi(\tau,\bs r) \,.
\label{eq:psiAction}
\end{equation} 

In Raman scattering, the interaction between the photons and the electrons takes the form $\rho(\tau,\vec{q})=\sum_{\vec{k},l,\sigma}\gamma_{\vec{k}} c^\dagger_{l,\vec{k}+\vec{q},\sigma}c_{l,\vec{k},\sigma}$ where $\sigma$ is the spin index and $\gamma_{\vec{k}}$ is the Raman vertex whose symmetry is related to the polarization of the incident and scattered photons. The Raman response function is related to the time-ordered correlation function $\chi_{\gamma\gamma}(\tau,\vec{q}=0) = -\langle T\rho(\tau,\vec{q}) \rho(0,-\vec{q}) \rangle$. To see how the photons couple to the collective mode, we introduce a source field $J$ which couples to $\rho$, $H_J(\tau) = \sum_{\vec{q}} J(\tau,\vec{q}) \rho(\tau,-\vec{q})$, which allows one to obtain $\chi_{\gamma\gamma}$ via a linear response theory about $J$. The Greens function perturbed with a small amplitude subleading pairing $\psi$ and $H_J$ becomes,
\begin{equation}
G^{-1} = G^{-1}_{0} + \Sigma \,,
\label{eq:newGreen}
\end{equation}
where $G_0$ is the unperturbed Gor'kov Green's function associated with the leading pairing $\Delta_0 f_{0\vec{k}}$, and the self energy is,
\begin{equation}
\Sigma = J\gamma_{\vec{k}}\sigma_3 + (\psi_1 \sigma_1 +\psi_2 \sigma_2)f_{1\vec{k}} \,,
\end{equation}
where $\psi_1$ and $\psi_2$ are the real and imaginary components of $\psi$. Combining (\ref{eq:psiAction}) and collecting nonvanishing terms in the perturbative expansion about $J$ up to the quadratic order, we arrive at the following effective action,
\begin{eqnarray}
S[\psi,J] &=& S[\psi]+ \int dq  \{  \sum_{i=1,2}[ J(-q)\Pi^\gamma_{J\psi,i}(q)\psi_i(q)   \nonumber \\
&+&   \psi_i(-q)\Pi^\gamma_{\psi J,i}(q) J(q) ] + J(-q)\Pi^{\gamma\gamma}_{JJ} J(q)  \}+ ... \,, \nonumber \\
~
\label{eq:EFFaction1}
\end{eqnarray}
where the correlation functions $\Pi^{\gamma\gamma}_{JJ}$ and $\Pi^\gamma_{J\psi,i}$ are given by,
\begin{eqnarray}
\Pi^l_{JJ}(q) &=&\frac{-1}{\beta}\sum_{k_0,\vec{k}} \text{Tr}[G_{0l}(k)\sigma_3G_{0l}(k+q)\sigma_3]\gamma^2_{\vec{k}} \nonumber \\
&\propto& \int dk \frac{\Delta_l^2\gamma^2_{\vec{k}}}{(k_0^2+E_{\vec{k}}^2)[(k_0+q_0)^2+E_{{\vec{k}}+{\vec{q}}}^2]}   \,.
\label{eq:Col2}
\end{eqnarray}
and 
\begin{eqnarray}
\Pi^\gamma_{J\psi,i}(q) &=& \Pi^\gamma_{\psi J,i}(q) =\frac{-1}{\beta}\sum_{k_0,\vec{k}} \text{Tr}[G_{0}(k)\sigma_i G_{0}(k+q)\sigma_3]\gamma_{\vec{k}}f_{1\vec{k}}
\label{eq:Col1}
\end{eqnarray}
which yields, 
\begin{eqnarray}
\Pi^\gamma_{J\psi,1}(q) &=& 0 \,,\, \text{and}\nonumber \\  
\Pi^\gamma_{J\psi,2}(q) &=& -\int dk \frac{2 q_0\Delta_{0}f_{0\vec{k}}f_{1\vec{k}} \gamma_{\vec{k}}}{(k_0^2+E_{\vec{k}}^2)[(k_0+q_0)^2+E_{{\vec{k}}+{\vec{q}}}^2]} \,.
\label{eq:Col1a}
\end{eqnarray}
We thus see that only the imaginary component of $\psi$ couples to the source field in (\ref{eq:EFFaction1}). Taking an isotropic superconducting gap for simplicity, in the limit relevant for Raman scattering $\vec{q}=0$ the $\vec{k}$-integration in the second equation of (\ref{eq:Col1a}) can be approximated by a Fermi surface integral, $\langle f_{0\vec{k}} f_{1\vec{k}} \gamma_{\vec{k}} \rangle_\text{FS}$. By symmetry, this immediately suggests the active Raman channel that is sensitive to the collective fluctuations of $\psi$: with $f_{0\vec{k}}$ and $f_{1\vec{k}}$ being $s$- and $d$-wave form factors (or vice versa), only the $B_{1g}$ Raman vertex $\gamma_{\vec{k}}^{B_{1g}} \propto \cos k_x - \cos k_y$ could ensure a nonvanishing $\Pi^\gamma_{J\psi,2}$ upon $\vec{k}$-integration! Hence the collective mode only manifests in this Raman channel. 

Integrating out $\psi$ in (\ref{eq:EFFaction1}), one arrives at an effective theory for the source field $J$: $S[J]=J(-q)\Pi_{JJ}(q)J(q)$, where
\begin{equation}
\Pi_{JJ}(q) = \Pi^{\gamma\gamma}_{JJ}(q) + \Pi^\gamma_{J\phi,2}(q) (q_0^2 +\frac{\bar{v}^2_F \vec{q}^2}{2}+w_L^2)^{-1}\Pi^\gamma_{\phi J,2}(q) \,.
\end{equation}
Setting $\vec{q}=0$, and making an analytic continuation $iq_0 \ra w + i0^+$, the imaginary part of $\Pi_{JJ}(w+0^+,\vec{q}=0)$ then returns the Raman response function we set out to derive. The first term results in a pair breaking peak at the continuum edge $w=2\Delta_0$, while the second term gives a $\delta$ peak at the Leggett (BS) mode frequency $w=w_L$. Note that as our primary purpose of the derivation is to show explicitly the symmetry aspect of the Raman response function, we have ignored vertex corrections which might lead to reduced and broadened peaks \cite{Maiti:17a}. These are beyond the scope of our study.  


\begin{thebibliography}{99}
\bibitem{footnote} In Leggett's original discussion, both leading and subleading states are $s$-wave pairing.
\end{thebibliography}
\end{document}